\documentclass[twocolumn]{aastex631}

\usepackage{graphicx}	
\usepackage{amsmath,ulem}	
\usepackage{natbib}
\usepackage{breqn}
\usepackage{amsmath}

\usepackage{bigints}
\usepackage{tikz}
\PassOptionsToPackage{dvipsnames}{xcolor}
\definecolor{webgreen}{rgb}{0,.5,0}

\definecolor{arsenic}{rgb}{0.23, 0.27, 0.29}

\begin{document}

\title{Between the cosmic-ray `knee' and the `ankle': Contribution from star clusters}

\author[0000-0002-0044-9751]{Sourav Bhadra}
\affiliation{Raman Research Institute, Sadashiva Nagar, Bangalore 560080, India}
\affiliation{Joint Astronomy Programme, Department of Physics, Indian Institute of Science, Bangalore 560012, India}

\author{Satyendra Thoudam}
\affiliation{Department of Physics, Khalifa University, PO Box 127788, Abu Dhabi, United Arab Emirates.}


\author{Biman B Nath}
\affiliation{Raman Research Institute, Sadashiva Nagar, Bangalore 560080, India}

\author{Prateek Sharma}
\affiliation{Joint Astronomy Programme, Department of Physics, Indian Institute of Science, Bangalore 560012, India}




\begin{abstract}
\setlength{\leftskip}{10em}
We show that massive young star clusters may be possible candidates that can accelerate  Galactic cosmic rays (CRs) in the range of $10^7\hbox{--}10^9$ GeV (between the `knee' and `ankle'). 
Various plausible scenarios such as acceleration at the wind termination shock (WTS), supernova shocks inside these young star clusters, etc. have been proposed, 
since it is difficult to accelerate particles up to the  $10^7\hbox{--}10^9$ GeV range in the standard paradigm of CR acceleration in supernova remnants. 
We consider a model for the production of different nuclei in CRs from massive stellar winds using the observed distribution of young star clusters in the Galactic plane. We present a detailed calculation of CR transport in the Galaxy, taking into account the effect of diffusion, interaction losses during propagation, and particle re-acceleration by old supernova remnants to determine the all-particle CR spectrum. 
Using the maximum energy estimate from the Hillas criterion, 
we argue that a young massive star cluster can accelerate protons up to a few tens of PeV. 
Upon comparison with the observed data, our model requires a CR source spectrum with an exponential cutoff of $5\times 10^7 Z$ GeV ($50\,Z$~PeV) from these clusters together with a cosmic-ray injection fraction of $\sim 5\%$ of the wind kinetic energy.
We discuss the possibility of achieving these requirements in star clusters, and the associated uncertainties, in the context of considering star clusters as the natural accelerator of the `second component' of Galactic cosmic rays.
\end{abstract}




\section{Introduction}
\label{sec:intro}
Cosmic rays (CRs hereafter) are high-energy particles that span an extensive range of energy from $1$ GeV to $\sim 10^{11}$ GeV. Lower energy CRs up to $\sim 10^{5-6}$ GeV are believed to be accelerated by supernova shocks (\citealt{Lagage1983}; \citealt{Axford1994}).  This dominant acceleration mechanism, revealed by both theoretical (\citealt{Fermi1949, Axford1977, Bell1978, Blandford1978, Blasi2013, Caprioli2015}) and observational \citep{Drury1994, Ackermann2013, Abdalla2018} studies, is diffusive shock acceleration (DSA), a first-order Fermi acceleration process in which $\sim 10$\% of the shock energy is expected to be converted to cosmic rays. Although the acceleration mechanism continues to work throughout the active stage of a supernova remnant (SNR) until it becomes indistinguishable from the ambient interstellar medium after $\sim 10^5-10^6$ years, most of the particle acceleration occurs during the un-decelerated blast wave phase, which lasts for $\le 10^3$ years 
\citep{Lagage1983}. This limits the maximum CR energy that can be accelerated in SNRs because the acceleration time of CRs cannot be longer than the age of the SNR \citep{Morlino2017}. Considering nonlinear effects such as the scattering of the cosmic rays by the waves they generate themselves and assuming the magnetic flux density of the interstellar magnetic field to be $\sim \mu$G, \citet{Lagage1983} estimated the upper limit of CR energy in supernova remnants to be $\sim 10^5$ GeV per nucleon. 

Preliminary observations seem to align with these theoretical concepts. \citet{Suzuki2022} reported cutoff energy of around TeV from $\gamma$-ray observations of $15$ supernova remnants. Recently, LHAASO (Large High Altitude Air Shower Observatory), and Tibet air shower observations have identified a number of PeVatron candidates (\citealt{Cao2021, Tibet2021}), which may include a few SNRs. These theoretical and observational developments suggest cutoff energy in the range $10^5-10^6$ GeV for SNRs. At the high-energy end, cosmic rays above $\sim 10^9$~GeV are considered to have an extragalactic origin, possibly originating from galaxy clusters \citep{Kang1996}, radio galaxies \citep{Rachen1993}, AGN jets \citep{Mannheim2000} or gamma-ray bursts \citep {Waxman1996}.

There is a gap between the contribution from SNRs and the extragalactic component, which lies in the range of $\sim 10^7-10^9$~GeV, the region between the so-called `knee' and `ankle' (also known as the `shin' region). To explain this gap in the all-particle CR spectrum, a few models have been proposed in the literature. \citet{Biermann1993, Thoudam2016} have discussed the explosion of Wolf-Rayet stars embedded in the wind material from the same stars as a potential acceleration site of CRs in the range of $\sim 10^7-10^9$ GeV. However, there may be some problems with this scenario. A uniform distribution of Wolf-Rayet stars in the Galaxy was assumed, which is unrealistic. Moreover, there are many uncertainties in the crucial parameter of the magnetic field of the Wolf-Rayet stars. For a proton cutoff energy of $1.1 \times 10^8$ GeV,  the surface magnetic field of a Wolf-Rayet star is required to be $\approx 10^4$ G in this model \citep{Thoudam2016}, while realistic predictions for the same are in the range of a few hundred G 
\citep{Neiner2015, Blazere2015}. 
Although no direct magnetic signature has been detected in any of the Wolf-Rayet stars, using Bayesian statistics, \citet{Bagnulo2020} have estimated their surface magnetic fields to be of the order of a few kiloGauss. 

The idea of Galactic wind termination shock to accelerate high-energy CRs also has problems. The effect of Galactic winds on the transport of cosmic rays in the Galaxy has been discussed in detail (\citealt{Lerche1982, Bloeman1993, Strong1998, Jones2001, Breitschwerdt2002}). Following these developments, \citet{Jokipi1987, Zirakashvili2006, Thoudam2016}   introduced these CRs originating from Galactic wind termination shock as the possible candidate for the `second' (between `knee' and `ankle') component of Galactic cosmic rays. The cosmic rays originating from the Galactic wind (GW-CRs) are believed to mostly contribute to the higher energy range. This is due to the increasing effect of advection over diffusion at lower energy, preventing particles from reaching the Galactic disk. Higher energy particles, which diffuse relatively faster, can overcome the advection and reach the disk more effectively. \citet{Thoudam2016} have used a distance of $\sim100$ kpc for the Galactic wind termination shock. \citet{Bustard2017} have argued that in order for the CRs to reach $10^8$ GeV, either the outflow speed needs to be of order $\sim 1000$ km s$^{-1}$ or the magnetic field needs to be amplified. However, realistic simulations of outflows from Milky Way-type galaxies do not find signatures of such strong outflows/shocks. \citet{Sarkar2015} showed that the outer shock due to the Galactic wind weakens and continues to propagate as a sound wave through the circum-galactic medium. 
The termination shock remains confined within $\lesssim 10$s of kpc and disappears after the mechanical power is stopped being injected. 
Also, the observed nuclear abundances suggest lighter nuclei in contrast to the expectation from the Galactic wind model in the $10^7-10^9$ GeV energy range. Thus, this model has been disfavoured.  
In order to explain the observed all-particle spectrum in the range $10^7-10^9$ GeV, an appropriate model of CRs is still required.

Coming back to the DSA mechanism of CR acceleration in supernova shocks, this standard scenario is known to bear several ailing problems (e.g., \citealt{Gabici2019}). The acceleration scenario cannot explain some of the observed features of cosmic rays like excess of Ne$^{22}$/Ne$^{20}$ in CRs compared to standard cosmic abundances in ISM (\citealt{Binns2008, Wiedenbeck1999}), proton acceleration up to greater than PeV ($10^6$ GeV) energy range, and so on. Various additional CR acceleration sites are reported in the literature to solve these paradigms; young massive star clusters are one of those other possible sources of Cosmic rays in our Galaxy \citep{Bykov2014, Knodlseder2013, Aharonian2019}. Recently, the $\gamma$-ray observations by LHAASO, HESS, Fermi-LAT, and HAWC have provided evidence of CR acceleration up to very high energy in a few massive star clusters like Westerlund1 and Cygnus \citep{Aharonian2019, Abeysekhara2021}. These star-forming regions have been discussed as potential candidates for CR accelerators (e.g. \citealt{Bykov2014} ); these $\gamma$-ray observations have strengthened the hypothesis of CR acceleration in these environments. Recently, \citet{Gupta2020} has shown that the excess ($^{22}$Ne/$^{20}$Ne) ratio can be explained by considering wind termination shock (WTS) of massive star clusters as CR accelerators. Recently, \citet{Tatischeff2021} showed that the refractory elements of Galactic cosmic rays are produced in super-bubbles. This theoretical and observational evidence prod us to study the total contribution of Cosmic rays originating from the distribution of massive star clusters in our Galaxy.

Star clusters, which are the birthplace of massive stars (that ultimately explode as SNe), 
give rise to continuous mass outflow in the form of stellar wind. These are mainly located in dense molecular clouds and weigh of the order of several thousand solar masses \citep{Longmore2014}. Star clusters host massive stars as well as supernova explosions, which produce a low-density bubble around them \citep{Weaver1977, Gupta2018}. Young star clusters contain sufficient kinetic energy supplied by interacting stellar winds, which can accelerate protons up to $\sim 10^7$ GeV. Considering heavier nuclei, this cosmic ray component originating from star clusters can, therefore, be considered as the second component of Galactic cosmic rays, which can explain the observed all-particle spectrum in the energy range of $10^{7}-10^{9}$ GeV. \citet{Bhadra2022}, using hydrodynamic simulation, showed that the observed distribution of $\gamma$-rays can be explained by invoking cosmic ray acceleration in the Westerlund1 cluster. 

Following the above discussion, it is clear that: (1) Galactic supernovae can accelerate particles up to a few times $10^6$ GeV energy, and (2) extragalactic components can explain the all-particle spectrum above $\sim 10^9$ GeV energy. The gap in the energy range cannot be explained using only these two components, and we require another Galactic component to explain the observed data in the range $10^7-10^9$ GeV. Our main focus in this paper is the second component of Galactic cosmic rays. In this regard, we propose CR contribution from the population of massive star clusters as a source of the observed all-particle CR spectrum in the range $\sim 10^7-10^9$ GeV. This may act as a bridge between the SNR component and the extragalactic component and fill the gap in the desired energy range.

We begin with some basics in Section \ref{sec:existing_components}. The details of our proposed model are described in Section \ref{sec:second_comp}. In Section \ref{sec:model_second},  we present our results, and in Section \ref{sec:all_particle} we compare our model with other models. In Section \ref{sec:vary_eg}, we consider various models for the extragalactic CR component. This is followed by further discussion in Section \ref{sec:discussion} and a conclusion in Section \ref{sec:conc}.

\section{Existing components of Cosmic rays}
\label{sec:existing_components}
\subsection{First Galactic component: SNR-CRs}
As mentioned in the Introduction, supernova remnants are the most likely candidate for cosmic-ray acceleration up to $\sim 10^6$ GeV energy \citep{Lagage1983}. The diffusive shock acceleration at strong shocks produces a power-law spectrum with an index of $\sim -2$ (\citealt{Krymskii1977, Bell1978, Blandford1978, Caprioli2011}). We have adopted the model of \citet{Thoudam2016} for the CR component from Galactic supernova remnants (SNR-CR component). After the acceleration in the strong shock of supernova remnants, CR particles escape the remnants and propagate through the interstellar medium via diffusion. These CR particles can be re-accelerated repetitively by expanding supernova remnant shock waves already existing in the interstellar medium during their propagation. These shocks are mainly produced by older remnants and are relatively weak. 

We use the same contribution of the SNR-CR component as presented in \citet{Thoudam2016}. Their calculation assumes an exponential cutoff for the proton source spectrum at $E_c=2.5\times 10^6$ GeV. This value has been chosen by fitting the observed all-particle spectrum. The maximum energy of SNR-CRs corresponds to the cutoff energy of iron nuclei, which is $26 \times E_c=6.5\times 10^7$ GeV. This result shows that SNR-CRs can contribute only $\sim 30\%$ of the total observed intensity above $\sim 2 \times 10^7$ GeV \citep{Thoudam2016}. Therefore, additional components are required to explain the all-particle spectrum in the $\gtrsim 10^7$ GeV range.

\subsection{Extragalactic component}
\label{section:2.2}
Various previous works have already pointed out that the `ankle'-like feature of the CR spectrum at $\gtrsim 10^9$ GeV can be explained if we consider the propagation effects of the extragalactic component (mainly proton) in the evolving microwave background \citep{Hillas1967, Berezinsky1988, Berezinsky2006, Aloisio2012, Aloisio2014}. 
We consider two different models for the extragalactic component: 
the `UFA model' \citep{Unger2015} and a combination of minimal (\citealt{dimatteo2015}), and PCS model (\citealt{Rachen2016, Thoudam2016}). We refer to this combined model as the `MPCS' model.

\citet{Unger2015} considers acceleration of energetic nuclei at the shocks associated with gamma-ray bursts or tidal disruption events, and photo-disintegration of these particles in the photon background present inside the source region. In this model, only the highest energy particles having an escape time shorter than the photo-disintegration time can escape the source region leading to a strong proton component in the energy region below the ankle. We call this the `UFA' model of the extragalactic component. In addition to the all-particle CR spectrum, data of the primary composition in the ultra-high energy range have become available in the last few years.

The `minimal model' has been derived from CR composition measured by the Pierre Auger Observatory (\citealt{dimatteo2015}) and assumes uniformly distributed sources in a comoving volume that produce a power-law cosmic ray spectrum with some cutoff at a particular rigidity $R_c$ (rigidity is defined as $Apc/Ze$, where $A/Z$ is nuclear mass/charge and $p$ is momentum, $e$ the charge of the electron, and $c$ is the speed of light in vacuum). Above $\sim 3 \times 10^{10}$ GeV, the spectrum exhibits a steep cut-off that is mainly due to the intrinsic cut-off in the injection spectrum \citep{dimatteo2015}, and not due to the GZK absorption \citep{Greisen1966, Zatsepin1966} during the propagation.

The PCS (primordial cluster shock) model is based on the universal scaling argument. It takes into account the acceleration of primordial proton and helium mixture by primordial cluster shocks, which are mainly the accretion shocks expected from clusters of galaxies during the structure formation. In this scenario, the acceleration of CR particles is limited by losses due to pair production in the CMB. This component is not expected to reach ultra-high energies. Consequently, the minimal model plus the “primordial cluster component” was introduced by \citet{Rachen2016}, where the acceleration of heavy nuclei at shocks of gamma-ray bursts or in tidal disruption events are considered.

\begin{figure}
\includegraphics[width=80mm
]{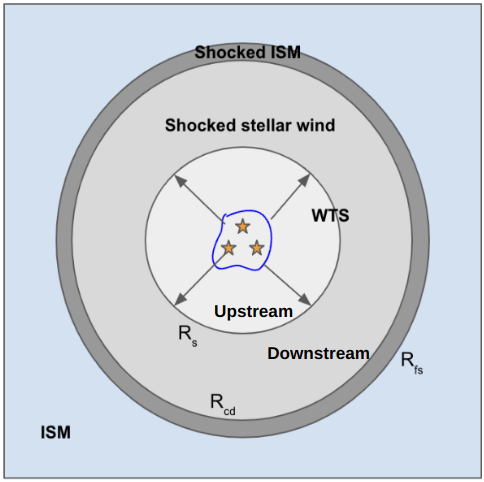}
\caption{Schematic diagram of a stellar wind bubble. The position of termination shock is $R_s$; $R_{\rm cd}$ and $R_{\rm fs}$ are contact discontinuity and forward shock positions, respectively.}
\label{fig:stellar_bubble}
\end{figure}

\section{Second Galactic component: cosmic rays from star clusters}
\label{sec:second_comp}
The all-particle cosmic ray spectrum has two main features: a steepening of the spectral index from $-2.7$ to $-3.1$ at about $3$ PeV, commonly known as the `knee', and a flattening back to $-2.7$ at about $4\times 10^9$ GeV, generally known as the `ankle'. 
Therefore, we need to assume a cut-off in the Galactic component immediately below the `ankle' to explain the observed spectrum. This is a `second knee' feature in the CR spectrum. For a typical magnetic field of $3$ $\mu$G in the Galaxy, cosmic rays with energy Z $\times 10^8$ GeV have a Larmor radius of $36/Z$ pc, which is much smaller than the extent of the diffusion halo of the Galaxy. This implies that cosmic rays with the energy around the second knee remain confined within the Galaxy. This also suggests the observed cut-off at this energy is due to some CR accelerators different from SNRs, as the latter accelerate particles only up to a few $10^6$ GeV. 

In the following, we discuss one potential scenario of another Galactic component of CRs: the acceleration of cosmic rays by the young massive star clusters, which we briefly mentioned in Section \ref{sec:intro}. It has especially been speculated that the winds of massive stars may be a suitable location for the acceleration of CRs (\citealt{Cesarsky1983, Webb1985, Gupta2018, Bykov2020}). CRs can be accelerated in the fast stellar wind of star clusters, and in particular, two scenarios can be important. Firstly, CR acceleration in the wind termination shock (WTS) \citep{Gupta2018}, and secondly, acceleration by supernova shocks embedded in the stellar winds \citep{Vieu2022}.  Those cosmic rays accelerated in young star clusters with age $\leq 10$ Myr can contribute significantly to the observed total flux of CRs \citep{Gupta2020}. Recently LHAASO has observed $\gamma$-rays in the PeV energy range from young massive star clusters \citep{Cao2021}, which can be associated with cosmic ray acceleration in those clusters. 

Figure \ref{fig:stellar_bubble} shows a schematic diagram of a stellar wind bubble around a compact star cluster. There are several distinct regions inside the bubble, such as (a) the free wind region, where the stellar wind originating from the source expands adiabatically, (b) the wind termination shock (WTS), (c) the shocked wind region containing slightly denser gas, and (d) the outermost dense shell containing the swept-up ambient gas. Cosmic rays can be accelerated in the central region as well as in the shocks of the cluster. After getting accelerated to very high energy, cosmic rays will diffuse outward from the source into the ISM.
\subsection{Distribution of star clusters in Galactic plane}
\label{section_3.2}
Star clusters are distributed all over the Galactic plane, and each star cluster creates a superbubble-like structure around itself  \citep{Weaver1977, Gupta2018}. 
\citet{Bronfman2000} observed $748$ OB associations across the Galactic disk and found their distribution to peak at $R_p = 0.55\,R_0$,\, ($R_0 = 8.5$ kpc is the solar distance from the Galactic center). We find that their inferred (differential) star cluster distribution can be roughly fitted by
\begin{equation}
    dN_c(r)=\Sigma_0\,e^{\frac{-(r-R_p)^2}{\sigma^2}}\, 2\pi r\,dr, 
\end{equation}
where $r$ is the Galactocentric distance and $\sigma=3$ kpc and $\Sigma_0$ is the normalization constant in the unit of kpc$^{-2}$.
This denotes the number of star clusters in an annular ring of radius $r$ to $r+dr$. The surface density of the clusters can be obtained by dividing this number by the surface area of the annular ring ($2 \pi r\,dr)$, as
\begin{eqnarray}
\label{eq:nu_clusters}
    \nu(r) = \Sigma_0\,e^{\frac{-(r-R_p)^2}{\sigma^2}}
\end{eqnarray}
where, $\Sigma_0 \sim 14$ kpc$^{-2}$ \citep{Nath2020}. We have used a minimum number of 10 and a maximum number of $1000$ OB stars in a cluster (these are somewhat arbitrary, and we later discuss the impact of these choices).

The actual distribution of cosmic ray sources is expected to follow the distribution of young stars and dense gas in the form of a spiral structure, for instance, as traced by the FIR luminosity \citep{Bronfman2000}, or the Ly-$\alpha$ radiation \citep{Higdon2013}, which show a bit of asymmetry and trace the spiral arms to some extent. We emphasize that although we assume an axisymmetric source distribution with smooth radial distribution, this assumption yields a spectrum for cosmic-ray protons/nuclei that closely resembles that derived from the spiral-arm feature of the source distribution (e.g., \citealp{Werner2015}). Beyond $\sim 10$ GeV, \cite{Werner2015} show
that the spectrum's shape remains largely unchanged and flux varies by less than 2\% when  spiral-arm features are introduced.
However, it is worth noting that the presence of nearby star clusters associated with spiral arms can introduce noticeable effects on cosmic-ray anisotropy at Earth. This is an 
important topic for future investigation, but 
beyond the scope of the current paper.
%
%
\subsection{Transport of CRs originating from star clusters in the Galaxy}
After getting accelerated in SNR and star cluster shocks, CRs propagate through the Galaxy. This propagation is mainly dominated by diffusion and energy loss due to interaction with ISM material. Some fraction of the propagating CRs can be re-accelerated up to higher energy by the interaction with existing weaker shocks that have been generated from older supernova remnants in the ISM.  This process has been discussed in detail in \cite{Thoudam2014}.
The transport equation for cosmic ray nuclei in a steady state can be written as
\begin{eqnarray}
\nabla\cdot(D\nabla N)-[n{\rm v}\sigma+\zeta]\delta(z)N+\nonumber\\ 
\Big[\zeta s p^{-s}\,\int_{p_0}^p\, du\, N(u)\,u^{s-1}\Big]\delta(z)
=-Q(r,p)\delta(z).
\label{transport_eqn}
\end{eqnarray}

Here we include spatial diffusion (first term on the left-hand side), re-acceleration (terms with coefficient $\zeta$), and interaction losses ($\propto \sigma$, the loss cross-section) of the CR particles, as mentioned above. The diffusion coefficient $D(p)$ depends on the momentum of CR particles.
Here $n$ represents the averaged surface density (number density per unit area) of interstellar atoms in the Galaxy,  $v(p)$ is the CR particle velocity, $\sigma(p)$ is the cross-section of inelastic collision, $N$ is the differential number density (number per unit volume per momentum) and $\zeta$ is the rate of re-acceleration. The third term involving the momentum integral represents the generation of higher energy particles via the re-acceleration of lower energy particles. It has been assumed that a CR population is instantaneously re-accelerated to form a power-law distribution with an index of $s \sim 4.5$ \citep{Thoudam2014}. We consider a cylindrical geometry for the diffusion halo denoted by the radial coordinate $r$ and vertical direction $z$. The diffusive halo has upper and lower boundaries at $z=\pm H$ and a radial boundary at $20$ kpc. A significant fraction of cosmic rays that reach the earth is produced from those sources located within a distance $\sim 5$ kpc \citep{Taillet2003}.  

The term on the right side, $Q(r, p)\delta(z$), represents the injection rate of cosmic rays per unit volume in the momentum bin $[p,p+dp]$ by the sources. The $\delta(z)$ term denotes that all sources are confined to the Galactic plane $z=0$. Similarly,  re-acceleration and loss regions are confined within the Galactic mid-plane. 

The injection term $Q(r, p)$ can be written as a combination of a space-dependent part and a momentum-dependent part, i.e., 
\begin{eqnarray*}
    Q(r, p) = \nu(r)\, H[R-r]\,H[p-p_0]\, Q(p),
\end{eqnarray*}
where $\nu(r)$ (see equation 2) represents the number of star clusters per unit surface area on the Galactic disk (see section \ref{section_3.2} for details), $H[t] = 1(0)$ for $t > 0(<0)$ is the Heaviside step function, and $p_0$ (which is the lower limit in the integral in Equation \ref{transport_eqn}) is the low-momentum cutoff introduced to approximate the ionization losses. \citet{Wandel1987} showed that the ionization effects could be taken into account if we truncate the particle distribution below $\sim 100$ MeV/nucleon. In our calculation, we introduce a low-energy cutoff of $100$ MeV/nucleon. Our assumed distribution of star clusters, motivated by observations, has a peak at $\sim 4.6$ kpc ($0.55 R_0$, where $R_0$ is the distance of the Earth from the Galactic center $\sim 8.5$ kpc) and then decreases rapidly at large distances (for details see section 3.1). 

The expression for surface density of star clusters $\nu$ has been calculated in section \ref{section_3.2}, and the power-law source spectrum is described in section \ref{section_3.3}. The energy-dependent diffusion coefficient as a function of particle rigidity follows
\begin{eqnarray*}
    D(\rho) = D_0\,\beta(\rho/\rho_0)^{\delta},
\end{eqnarray*}
where $D_0$ is the diffusion constant, $\rho = Apc/Ze$ is the particle rigidity, $\beta = v(p)/c$ where $v(p)$ is the CR particle velocity and $c$ is the speed of light, $\delta=0.33$ is the diffusion index, and $\rho_0=3$ GV is a normalisation constant.

In this injection-diffusion-reacceleration\footnote{For typical parameters, reacceleration only affects the CR spectrum below $10^5$ GeV \citep{Thoudam2014} and so is irrelevant for the energy range considered in this work.} model, the rate of reacceleration depends on the rate of supernova explosions and the fractional volume occupied by SNRs in the Galaxy. The reacceleration parameter $\zeta$ can be expressed as, $\zeta=\eta V\nu_{SN}$, where $V = 4\pi\mathfrak{R}^3/3$ is the volume occupied by each SNR of radius $\mathfrak{R}$ re-accelerating the cosmic rays. Here, $\eta$ is a correction factor that takes care of the actual unknown size of the remnants, and $\nu_{SN}$ is the rate of supernova explosions per unit surface area in the Galactic disk. The values of $\mathfrak{R}$ and $\nu_{SN}$ have been taken as $100$ pc and $25$ SNe Myr$^{-1} $kpc$^{-2}$ respectively \citep{Thoudam2014}. 

\begin{figure}
\includegraphics[width=85mm
]{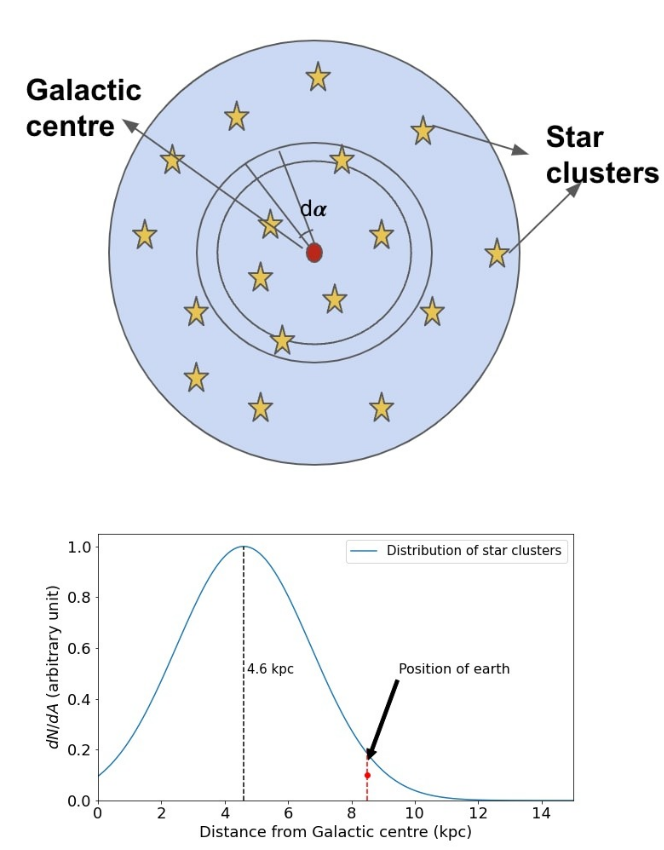}
\caption{Top: Schematic distribution of star clusters in the Galactic plane (face-on view), each star indicates a star cluster on the plane; bottom: the surface density (number per area) of star clusters ($\Sigma$; see Eq. \ref{eq:nu_clusters}) as a function of distance from the Galactic center.} 
\label{fig:dist_cluster}
\end{figure}

The solution of equation \ref{transport_eqn} can be obtained by invoking the Green's function method and by considering the two separate transport equations for the regions below and above the Galactic disk ($z < 0$ and $z > 0$ respectively), and by connecting the two solutions at Galactic plane, i.e., $z = 0$, via a jump condition. Following this procedure one can get the Green's function $G(r,r^{\prime},z,p,p^{\prime})$ (equation A.20 of \citealt{Thoudam2014}). After convolving the obtained Green's function with the assumed source distribution and integrating it over the Galactic plane, one can get the final solution (see equation 6 of the same paper) for the CR density $N(r,z,p)$. Following the procedure described in \citet{Thoudam2014}, we get the solution of the transport equation \ref{transport_eqn} as

\begin{eqnarray*}
    N(r,z,p) = 2\pi\,\int_{0}^{\infty}dp^{\prime}\int_{0}^{\infty} r^{\prime}dr^{\prime}\,G(r,r^{\prime},z,p,p^{\prime})\, Q(r^{\prime},p^{\prime}).
\end{eqnarray*}

Substituting the obtained $G(r,r^{\prime},z,p,p^{\prime})$ \citep{Thoudam2014} and the assumed source distribution in the above equation, the cosmic ray density at the Earth ($r= 8.5$ kpc) can be calculated by evaluating the above solution at $z = 0$ since our Solar system lies close to the Galactic plane. More explicitly, the differential number density measured at the location of Earth is
\begin{eqnarray}
N(r,p) &=& \int_{r^{\prime}=0}^{R}\,\int_{k=0}^{\infty} \Sigma_0\frac{J_0[k(r-r^{\prime})]}{L(p)}\,k\,dk\,e^{\frac{-(r^{\prime}-R_p)^2}{\sigma^2}}\,r^{\prime}\,dr^{\prime}\nonumber\\
&& \Big[Q(p) +\zeta s p^{-s}\,\int_{p_0}^p\, {p^{\prime}}^s\,dp^{\prime}\,Q(p^{\prime})\,A(p^{\prime})\,\nonumber\\
&& \times \exp \Big(\zeta s\int_{p^{\prime}}^p\, A(u)du\Big)\Big],\nonumber\\
\label{eq:number_density}
\end{eqnarray}
where $R=20$ kpc is the radial boundary of the Galaxy, $\Sigma_0$ is the number density of star clusters, $J_0$ is the Bessel function of order zero and the functions $A(p)$ and $L(p)$ are given by (see \citealt{Thoudam2014} for details),
\begin{eqnarray}
L(p) &=& 2D(\rho)k\, \coth (kH) + nv\sigma(p)+ \zeta\,,\\
A(p)&=&\frac{1}{pL(P)}.
\end{eqnarray}
Equation \ref{eq:number_density} gives the differential number density, i.e., number per unit volume per unit momentum of cosmic ray particles measured at earth. All the necessary terms needed to solve equation  \ref{eq:number_density} are discussed in sections 3.2--3.5.

\subsection{Injection spectra of cosmic ray nuclei}
\label{section_3.3}

The cosmic ray source spectrum $Q(p)$ from star clusters is assumed to follow a power-law in total momentum $Ap$, with an exponential cut-off, where $A$ is the mass number of the nucleus. We write the differential number of CR particles with nucleon number $A$, having momentum per nucleon in the range $(p\,, p+dp)$, as,  
\begin{equation}
    Q(p)= Q_0(Ap)^{-q} \exp\left(-\frac{Ap}{Zp_{\rm max}}\right)\,.
    \label{eq:injection}
\end{equation}
Here $Q_0$ is a normalization constant that is proportional to the fraction of total wind kinetic energy $f$ channeled into cosmic rays by a single star cluster. We call this `injection fraction', which is a free parameter and can be estimated by comparing the model result with observations. Also, $q$ is the spectral index, $p_{\rm max}$ is the cutoff momentum (for a single nucleon), and $Ap$ is the total momentum of a particle with the mass number $A$ and the atomic number $Z$.
\subsection{Maximum energy estimate of accelerated particles}
\label{section_max_energy}
For the estimation of the maximum accelerated energy of cosmic ray particles, we consider two different acceleration scenarios inside a young star cluster: acceleration at WTS and acceleration of particles around SNR shock inside a star cluster.
\subsubsection{Acceleration at wind termination shock (WTS):}
In equation \ref{eq:injection}, $p_{\rm max}$, which represents the maximum momentum of accelerated CRs, depends on the extension of the accelerating region for a stellar wind bubble of the cluster. Typically this accelerating region can be taken as the distance to the WTS ($R_{\rm WTS}$) from the center of the cluster. The maximum energy is achieved when the diffusion length becomes comparable to the size of the shock (in this case the WTS), for beyond this limit, the particles escape out of the accelerating region. In the case of Bohm diffusion, $\kappa = pc^2 /( \zeta qB )$, the maximum energy is then \citep{Vieu2022}:
\begin{eqnarray}
    E_{\rm max} \sim \zeta\, q\, B_{\rm WTS}\, {R_{\rm WTS}}\, \frac{V_w} {c} \,.
    \label{eq:hilas_criteria}
\end{eqnarray}

Here $R_{\rm WTS}$ is the radius of WTS. In the above equation, $V_w$ is the velocity of stellar wind, $B_{\rm WTS}$ is the value of the magnetic field at the WTS position, $\zeta = 3 r_g / \lambda$, with $\lambda$ the mean free path due to the magnetic field. The Bohm diffusion, which is the most optimistic scenario, corresponds to the limit $\zeta = 3$.

We follow the arguments advocated by \citet{Vieu2022} to estimate the magnetic field in the cluster core,
\begin{eqnarray}
    B_c \sim 150\left( \frac{n_{\rm c}}{10\,{\rm cm^{-3}}}\right)^{1/6}\left(\frac{\eta_T}{0.1}\right)^{1/3}
    \left(\frac{N_{\rm OB}}{100}\right)^{2/9}\nonumber\\ \left(\frac{R_c}{1 {\rm pc}}\right)^{-2/3}\,\mu G \,.
\end{eqnarray}
Here, $n_{\rm c}$ is the core density, $\eta_T$ is the efficiency of generation of turbulence, $N_{\rm OB}$ is the number of OB stars in the cluster, $R_c$ is the core radius of the cluster. The magnetic field advected into the free wind region has a $1/ r$ radial profile. Therefore, the magnetic field at the position of the wind termination shock and cluster core can be related using $B_cR_c=B_{\rm WTS}R_{\rm WTS}$. Therefore, equation \ref{eq:hilas_criteria} can be expressed as, 
\begin{eqnarray}
    E_{\rm max} \sim \zeta\, q\, B_{c}\, {R_{c}}\, \frac{V_w} {c} \,.
    \label{eq:hilas_criteria_2}
\end{eqnarray}
This leads to a maximum estimate:
\begin{eqnarray}
     E_{\rm max} \sim 6 \left(\frac{\zeta}{3}\right)\left( \frac{n_{\rm c}}{10\, {\rm cm^{-3}}}\right)^{1/6}\left(\frac{\eta_T}{0.2}\right)^{1/3}\left(\frac{N_{\rm OB}}{1000}\right)^{2/9}\nonumber\\
     \left(\frac{R_c}{1~{ \rm pc}}\right)^{1/3}\, \left(\frac{v_w}{2000 \, {\rm km\,s^{-1}}}\right) \,{\rm PeV}\nonumber\\.
     \label{eq:E_max}
\end{eqnarray}
Equation \ref{eq:E_max} gives a conservative estimate of ${\rm E_{max}=6\, PeV}$ ($6\times 10^6$ GeV) for the maximum attainable energy of protons. Note that this value is a few times higher than the maximum accelerated energy for the SNR-CR scenario.

However, in the realistic scenario, the magnetic field may be amplified in the accelerating region due to the existence of turbulence, and due to instabilities driven by cosmic ray streaming in the upstream region of WTS, which can therefore increase the estimated value of maximum accelerated energy. There are other uncertainties as well (e.g., in $\eta_T$, wind velocity $v_w$) 
that can conceivably increase the maximum energy by a factor of a few.
\subsubsection{Acceleration at SNR shock inside star clusters:}
Another potential scenario for CR acceleration inside the young star cluster is the SNR shocks propagating in the free wind region of the cluster. CR particles can be accelerated up to $10^8$ GeV if the SNR shocks advance through fast and highly magnetised stellar winds \citep{Volk1988, Biermann1993}. 
Non-linear effects in the acceleration process \citep{Bell2001} also contribute to this high-energy acceleration. \citet{Bell2013, Schure2013} 
highlight that the outer shocks of SNR can accelerate CR beyond the `knee' if the shock propagates into a magnetic field much
larger than a typical interstellar field, that can be present inside a star cluster.
Particles will be accelerated during the expansion of the SNR shock in upstream of the WTS. This idea has been studied extensively by \cite{Vieu2022}, and the maximum energy has been estimated by the authors as follows:
\begin{eqnarray}
    E_{\rm max} \sim 21\, \left(\frac{V_c}{5000 \,{\rm km\,s^{-1}}}\right)\,\left(\frac{\zeta}{3}\right)\,\left(\frac{R_c}{1\,{\rm pc}}\right)\,\left(\frac{N_{\rm OB}}{1000}\right)^{2/9}\,\nonumber\\
    \left( \frac{n_{\rm c}}{10\, {\rm cm^{-3}}}\right)^{1/6}
    \left(\frac{\eta_T}{0.2}\right)^{1/3}\, \left[1-\left(\frac{R_c}{R_{\rm WTS}}\right)^{1/7}\right]\,\,\,\, {\rm PeV}\,\,.\nonumber \\
\end{eqnarray}

For a typical young cluster $R_{\rm WTS}/R_c \sim 5-30$, which gives $\left(1-(R_c/R_{\rm WTS})^{1/7}\right) \sim 0.2-0.4$ \citep{Vieu2022}. This estimate can give a maximum energy of a few PeV for protons. However, if a supernova launches a very fast shock in the free wind region of a compact cluster with velocity $\geq 2\times 10^4$ km s$^{-1}$, it can accelerate protons up to a few tens of PeV energy. Note that, this high velocity of SNR shock inside a clumpy star cluster may efficiently drive MHD turbulence to generate a high value of the magnetic field, which will likely result in a higher value of maximum energy.
\begin{figure*}
\includegraphics[width=\textwidth, height=11cm]{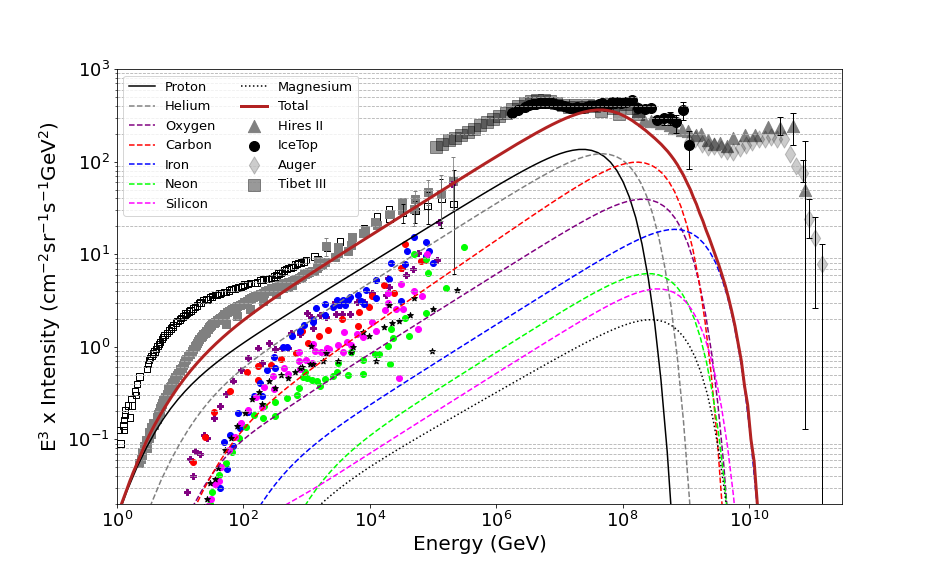}
\caption{Model prediction for the star cluster model as a second galactic component considering an injection fraction $\sim 5\%$. The thick solid maroon line represents the total contribution from  Galactic star clusters. Thin dashed lines represent the flux of individual elements. For the CRs generated from star clusters, an exponential energy cut-off for protons at $E_c = 5 \times 10^7$ GeV ($50$ PeV) is assumed.  High-energy data:
 IceTop \citep{Aartsen2013}, Tibet III \citep{Amenomori2008}, the Pierre Auger Observatory \citep{2013arXiv1307.5059T}, and HiRes II \citep{Abbasi2009}. Low energy data have been taken from CREAM \citep{Ahn2009, Yoon2011}, ATIC-2 \citep{Panov2007}, AMS-02 \citep{Aguilar2015a}, PAMELA \citep{Adriani2011}, CRN \citep{Mueller1991}, HEAO \citep{Engelmann1990}, TRACER \citep{Obermeier2011}, KASCADE \citep{Antoni2005}, DAMPE \citep{An2019}. We have only shown the high-energy data points with different symbols in the figure. Low data points: Proton (black square), Helium (grey square), Oxygen (purple solid plus), Carbon (red circle), Iron (blue circle), Neon (green circle), Silicon (magenta circle), Magnesium (black stars). The lower energy data from various experiments are represented together by one symbol. The error bars for proton and helium have been shown and the rest are not shown in the figure.}
\label{fig:cluster_comp}
\end{figure*}

The recent detection of $\gamma$-rays above PeV by {\it LHAASO} from some sources indeed indicates these sources can accelerate particles up to at least a few tens of PeV because, at high energy, the $\gamma$-ray energy can be approximated as $E_{\rm cr} \approx 10 E_{\gamma}$. Some of those sources possibly are young massive clusters (see extended table 2 of \citep{Cao2021}. The $\gamma$-ray photons from the LHAASO J2032+4102 source have the highest energy of 1.4 PeV, which corresponds to tens of PeV for cosmic ray protons. 
\subsection{Elemental abundances in star cluster winds}
\label{Section_3.4}
We consider a simple stellar population formed at time $t=0$ with an initial mass function (IMF) $\frac{dn}{dm}\propto m^{-2.35}$ \citep{Salpeter1955}. We can calculate the elemental abundances in the wind material following the procedure described in \citet{Roy2021}. Now,
\begin{equation}
    M_w(X,m,t)=\int_{0}^t\,\dot{m}_w(X,m,t^{\prime})\,dt^{\prime}
\end{equation}
is the cumulative mass of element $X$ ejected in winds by a star of initial mass $m$ up to age $t$ where,
\begin{eqnarray}
    \dot{m}_w(X,m,t^{\prime}) &=& {\rm mass \,\,\,fraction}(X,m,t^{\prime}) \times \dot{m}(m,t^{\prime})\nonumber\\
    &=& f(X,m,t^{\prime}) \times \frac{dm_{\rm star}}{dt^{\prime}}.
\end{eqnarray}
We use the mass loss rate for each nucleus $\dot{m}(X,m,t^{\prime})$ using models for nucleosynthesis in massive stars and their return to the ISM via winds (A. Roy, private communication). Hence the elemental abundance of a particular element $X$ can be calculated using,
\begin{eqnarray}
f(X,m)=\frac{M_w(X,m,t)}{M_{w,\rm tot}(m,t)} =\frac{\int_{0}^t\,\dot{m}_w(X,m,t^{\prime})\,dt^{\prime}}{\int_{0}^t\,\dot{m}(m,t^{\prime})\,dt^{\prime}}.
\end{eqnarray}
We have taken into account evolution until the core carbon burning time, which implies the maximum time of the evolution of a star with mass $m$ as the upper limit of the integration. The mass-weighted elemental abundance of element X can be calculated using the following expression invoking the Salpeter mass function,
\begin{eqnarray}
\langle f(X) \rangle = \frac{\int_0^m\,f(X,m)\,{Am^{-2.35}} dm}{\int_0^m\,{Am^{-2.35}} dm}
\end{eqnarray}
Using this method, we have calculated the mass-weighted mean individual elemental abundance in the ejected stellar wind material. We have used the results of a state-of-the-art evolutionary model. The elemental abundances have been calculated considering the rotation-driven instabilities inside the star, the correct abundances of elements, and the mass loss rate from the stellar surface.
\begin{figure*}
\includegraphics[width=\textwidth, height=11cm]{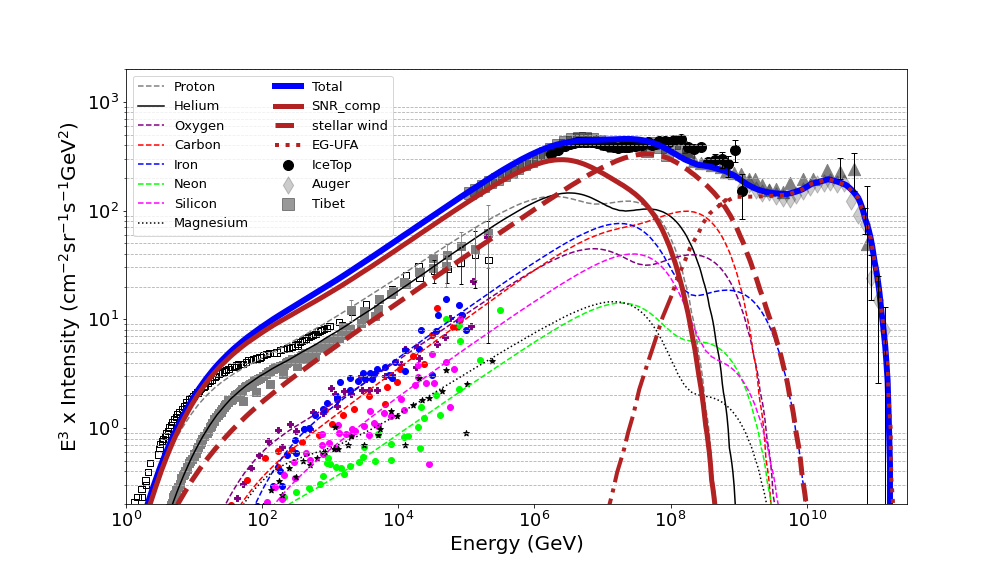}
\caption{Model prediction for the all-particle spectrum using the Galactic star cluster CR model as the second galactic component. For the star cluster component, the considered injection fraction is $\sim 5\%$, and the cutoff is at $5\times 10^7 Z$ GeV. The thick solid maroon line represents the total SNR-CRs, the thick dashed maroon line represents star cluster CRs, and the thick maroon dotted line represents the UFA model of extragalactic CR component (EG-UFA) taken from \citet{Unger2015}, 
and the thick solid blue line represents the total all-particle spectrum. The thin lines represent the total spectra for the individual elements i.e., a combination of both SNR-CR and the CRs originating from star clusters. The figure shows the $E^3$ times the cosmic ray flux $I(E) = (c/4\pi)N(E)$ at the position of the earth measured by different experiments as a function of cosmic ray energy, where $N(E)$ is the differential number density of cosmic ray particles. High energy and low energy data are the same as figure \ref{fig:cluster_comp}.}
\label{fig:all_comp}
\end{figure*}
\subsection{Average kinetic luminosity of clusters:}
Our assumption requires a certain fraction of the total wind kinetic energy to go into CRs. Therefore, we need the value of the average kinetic luminosity of a cluster using a distribution of OB associations over the luminosity range. \citet{Oey1997} assume that the mechanical luminosity function of OB association is given by $\phi(L) \propto L^{-2}$. We use this distribution to calculate the average luminosity of clusters with kinetic luminosity in the range $L_{\rm min}=10^{37}$ erg s$^{-1}$ (corresponds to $N_{\rm OB}=10$) to $L_{\rm max}=10^{39}$ erg s$^{-1}$ (corresponds to $N_{\rm OB}=1000$) as following,
\begin{eqnarray}
\langle L_w \rangle&=&\frac{\int_{L_{\rm min}}^{L_{\rm max}}\phi(L)\,L\,dL}{\int_{L_{\rm min}}^{L_{\rm max}}\phi(L)\,dL}\sim 4.5 \times 10^{37} \,\, {\rm erg\, s^{-1}}.
\end{eqnarray}
Note, we adopt a minimum $10$ number of OB stars for the production of CRs, and the largest OB association in our Galaxy has $1000$ OB stars. The dependence of $\langle L_w \rangle$ on $L_{\rm max}$ is weak, but there is the sensitive dependence on $L_{\rm min}$, the implications of which we discuss later.
\begin{figure*}
\includegraphics[width=\textwidth, height=10cm]{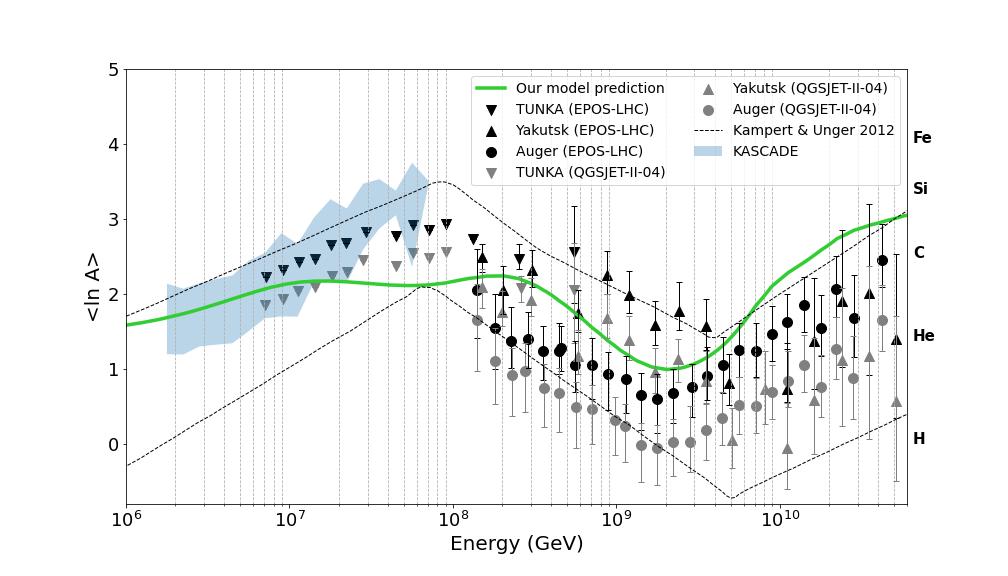}
\caption{Mean logarithmic mass $\langle \ln A \rangle$ of cosmic rays as a function of energy,  predicted using the combination of SNR-CR, CRs from star clusters (these two are Galactic components), and EG-UFA model (extragalactic component, \citealt{Unger2015}). Data:  KASCADE (\citealt{Antoni2005}), TUNKA (\citealt{Bereznev2013}), Yakutsk (\citealt{Knurenko2011}), the {\it{Pierre Auger}} observatory (\citealt{2015arXiv150903732T}) and the different optical measurement compiled in \citealt{Kampert2012}. The two different colored (black and grey) sets of data points correspond to two models EPOS-LHC and QGSJET-II-04, respectively, which have been used to convert $X_{\rm max}$ values to ${\langle \ln A \rangle}$ (see equation \ref{eq:lnAi}).} 
\label{fig:log_mass}
\end{figure*}
\section{Model prediction for the second component of galactic cosmic rays}
\label{sec:model_second}
The values of cosmic ray propagation parameters ($D_0,\,\delta$; the normalization of the diffusion coefficient and its power-law dependence on momentum) and re-acceleration parameters ($\eta\,, s$; the SNR filling factor and reacceleration power-law index) have been calculated by comparing the observed Boron to Carbon abundance ratio with the value obtained by the adopted model. The best fit values are $D_0=9 \times10^{28}$ cm$^2$s$^{-1}$, $\eta=1.02$, $s=4.5$, and $\delta=0.33$ \citep{Thoudam2014}. We have also used these values in our model. For the interstellar matter density ($n$), the averaged density in the Galactic disk within a radius equal to the size of the diffusion halo $H$ was considered. We choose $H = 5$ kpc \citep{Thoudam2016}, which gives an averaged surface density of atomic hydrogen of
$n = 7.24 \times 10^{20}$ atoms cm$^{-2}$ \citep{Thoudam2014}. To account for the helium abundance in the interstellar medium, we add an extra $10$\% to $n$. The radial extent of the source distribution is taken as $R = 20$ kpc. The inelastic cross-section for proton ($\sigma(p)$) is taken from \citet{Kelner2006}. 

Since we are interested in the acceleration of CRs in WTS, as well as around SNR shocks inside the free wind region of the star cluster, the relevant abundances correspond to that in the stellar wind for massive stars. 
For this purpose, we use the stellar wind abundances for massive stars beginning with the Zero Age Main Sequence (ZAMS) phase. We have used the surface abundance of massive stars as a function of time, calculated after properly taking into account the effect of stellar rotation. The spectral indices for different elements are given in Table 1. Note that these values are slightly different from the spectral indices assumed in \citet{Thoudam2016} for the SNR-CR component.
Also, the stellar wind elemental abundances are mentioned in Table 1, which are calculated 
using the method described in \cite{Roy2021} (provided to us by A. Roy, private communication). We have then averaged the abundances over time and mass distribution of stars in the cluster, as described in section \ref{Section_3.4}. Using these values of various parameters, we calculate the particle spectra for different cosmic ray elements (proton, helium, carbon, oxygen, neon, magnesium, silicon, and iron). The CR spectral indices ($q$) of source spectra for the individual elements are very similar to each other and are chosen to match the observed individual nuclear abundances in CRs closely.
\begin{table}
\centering
 \begin{tabular}{|c| c|c|} 
 \hline\textbf{Elements} & \textbf{$q$} & \textbf{Fractional abundances in winds}\\
 \hline \hline
 Proton & 2.25 & 0.86\\\hline
 Helium & 2.23 & 0.13\\ \hline
 Carbon &  2.20 & $3.32\times10^{-3}$ \\ \hline
 Oxygen &  2.24 & $8.51\times10^{-4}$ \\ \hline
 Neon &  2.24 & $8.83\times10^{-5}$ \\ \hline
 Magnesium &  2.28 & $3.62\times10^{-5}$ \\ \hline
 Silicon &  2.24 & $3.42\times10^{-5}$ \\ \hline
 Iron &  2.24 & $3.72\times10^{-5}$ \\ \hline
\end{tabular}
\label{Table:1}
\caption{Source spectral indices $q$ and fractional abundances of different elements in the wind material. The elemental abundances are calculated following \citet{Roy2021}.}
\end{table}\\

Figure \ref{fig:cluster_comp} shows the star cluster contribution to CRs using different parameters mentioned earlier. We have used the maximum energy for the proton as $5\times10^7$ GeV ($50$ PeV) and the injection fraction of $\sim 5\%$. These values of the parameters are chosen to match the observed spectra with our theoretical model. 
It is important to mention that, in section \ref{section_max_energy} we have estimated the maximum accelerated energy considering different scenarios in a star cluster. The maximum energy can go up to a few tens of PeV (especially for the SNR shock inside the star cluster scenario), although our used value is admittedly on the higher side. Also, recently \cite{Vieu2023} have shown that the SNR shocks inside a star cluster scenario can explain the all-particle cosmic ray spectrum in the region between `knee' and `ankle'. Therefore, star clusters are likely a possible candidate for cosmic ray acceleration between a few times $10^6$ and $10^9$ GeV.

Also, if one uses a higher lower limit of $N_{\rm OB}=30$ instead of $10$, then the injection fraction will need to be increased to match the observed spectrum. The data points correspond to different measurements. For lower energy ranges, the individual spectra are fitted to the observed elemental spectra. We consider $8$ elements: proton, helium, carbon, oxygen, neon, magnesium, silicon, and iron for our calculations, and the total contribution (solid brown curve in the figure \ref{fig:cluster_comp})  is a combination of these $8$ elements.
\section{All-particle spectrum of cosmic rays}
\label{sec:all_particle}
 
Figure \ref{fig:all_comp} combines all three CR components to get the total all-particle spectrum of cosmic rays and compares it with various observations.
The SNR-CR component shown in this figure is calculated following the procedure mentioned in \citet{Thoudam2016}, assuming a uniform distribution of SNRs in the Galactic plane and a proton spectrum cut-off of $\sim 2.5\times 10^6$ GeV.
For the extragalactic component, we have adopted the \textit{UFA} model \citep{Unger2015}, which considers a significant contribution of extragalactic CRs below the ankle to reproduce the observed CR energy spectrum as well as $X_{\rm max}$ (the depth of the shower maximum) and the variance of $X_{\rm max}$ above the ankle observed at the Pierre Auger Observatory \citep{dimatteo2015}. With these two models (SNR \& extragalactic), we have combined our proposed star cluster model with a proton spectrum cut-off at $5\times 10^7$ GeV ($50$ PeV).

The total contributions from all these three components can explain the observed features in the all-particle spectrum. Also, the spectra of the individual elements can be explained well with the model. The flux of different elements has been measured well in the lower energy region, but in the higher energy range, i.e., above $10^{5-6}$ GeV, the observation data for individual elements are not available. Observed data points for all-particle CR spectra have been taken from various experiments like TIBET III \citep{Amenomori2008}, IceTop \citep{Aartsen2013}, Auger \citep{2013arXiv1307.5059T}, HiRes II \citep{Abbasi2009}, etc.

Several ground-based experiments such as KASCADE, TUNKA, LOFAR, and the Pierre Auger Observatory have provided measurements of the composition of CRs at energies above $\sim 10^6$ GeV. Heavier nuclei interact at a higher altitude in the atmosphere, which results in smaller values of $X_{\rm max}$ as compared to lighter nuclei. For comparison with the theoretical predictions, $ \langle \ln A \rangle$, the mean logarithmic mass of the measured cosmic rays, is of utmost importance. This can be obtained from the measured $X_{\rm max}$ values using the following relation mentioned in \citet{Horandel2003},
\begin{eqnarray}
\label{eq:lnAi}
{\ln \, A}_i \,=\, \Big(\frac{X_{\rm max}^i-X_{\rm max}^{\rm p}}{X_{\rm max}^{\rm Fe}-X_{\rm max}^{\rm p}}\Big)\,\times {\ln \,A_{Fe}}.
\end{eqnarray}
Here $X_{\rm max}^{\rm p}$ and $X_{\rm max}^{\rm Fe}$ represent the average maximum depths of the shower for protons and iron nuclei, respectively, and ${\rm A_{Fe}}$ is the mass number of iron nuclei.
In figure \ref{fig:log_mass}, we have also shown the obtained mean logarithmic mass using our model and compared it with the observational data.

We calculate the mean mass in the following way,
\begin{eqnarray}
\langle {\ln\, A} \rangle=\frac{{\sum_{i}\ln \, A_i}\times {\rm Flux_i}}{\sum_{i}\,{\rm Flux_i}}
\end{eqnarray}
where ${A_i}$ denotes the mass number of an element $i$ (we have considered $8$ elements: proton, helium, carbon, oxygen, neon, magnesium, silicon, and iron), and ${\rm Flux_i}$ is the obtained flux of element $i$ using our model. 
Figure \ref{fig:log_mass}  shows that the results obtained using our star cluster model (green curve) follow the observed trend for the mean logarithmic mass in the total energy range from $10^8$ GeV to $10^{11}$ GeV when combined with the UFA model for the extragalactic CRs. In the energy range of about $2\times10^7$ and $10^8$ GeV, our prediction shows some deviation from the observed trend but still lies within limits presented in  \citet{Kampert2012}.

To reiterate, the primary focus of this work has been to present a model incorporating stellar wind shocks that can explain the observed all-particle CR spectrum, especially in the energy range between $10^7$ and $10^9$ GeV. The discussion so far shows that we can indeed explain the observed data in this energy range using a CR component originating from massive star clusters. The required CR injection fraction for this component of $\sim 5\%$ and an energy cutoff of $ 5\times 10^7 \, Z$ GeV ($50$ PeV),
as suggested by the fitting of the all-particle CR spectrum with our proposed stellar wind model,  are entirely reasonable, and therefore lend support to the idea that CR from massive star clusters can fill the CR spectrum gap between the `knee' and the `ankle'. We will discuss the implications of our result in Section \ref{sec:discussion}. Before that, we discuss the dependence of the required injection fraction and cutoff energy on the chosen extragalactic component in section \ref{sec:vary_eg} below. 
\section{Varying the extragalactic component}
\label{sec:vary_eg}
\begin{figure}
\includegraphics[width=92mm]{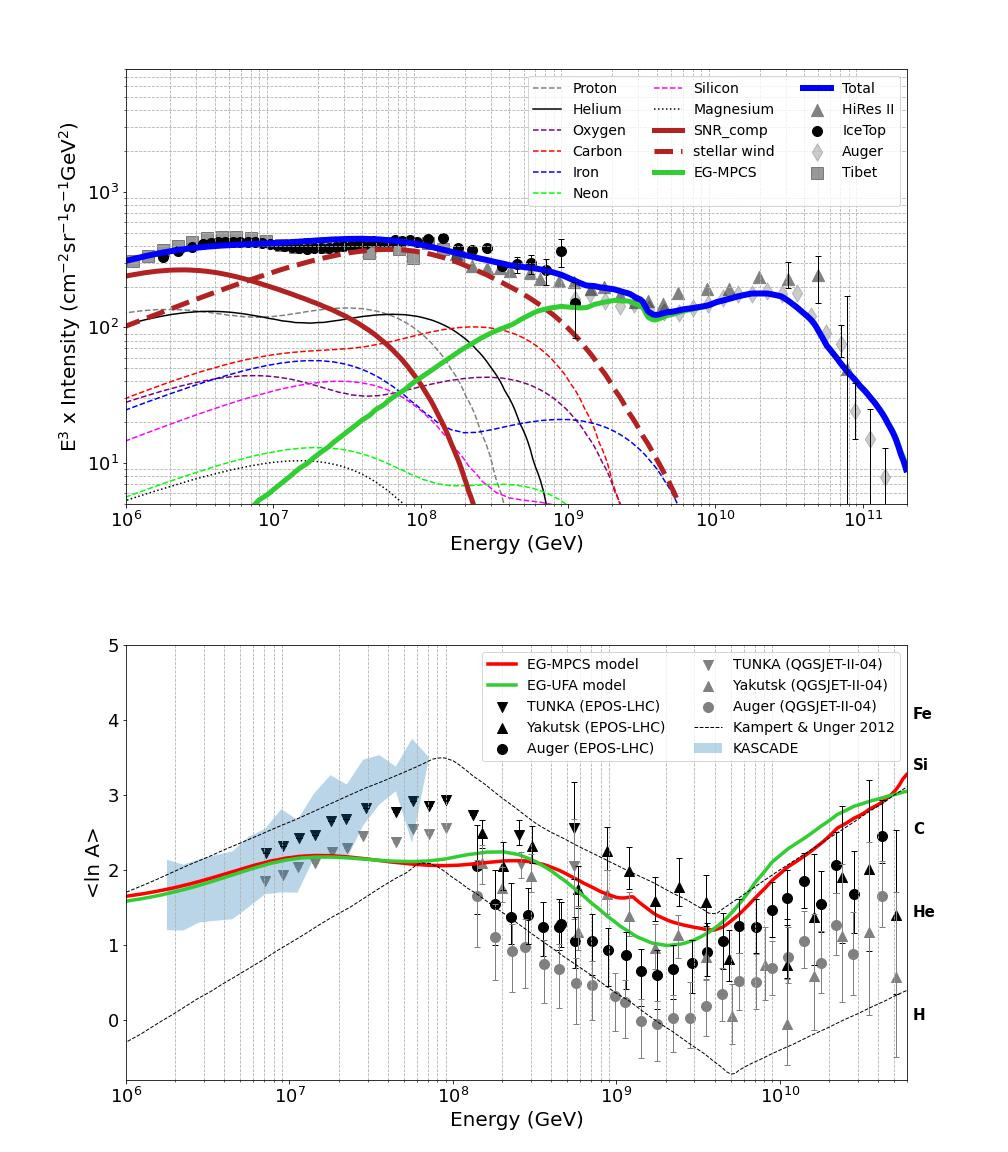}
\caption{\textit{Top panel:} All-particle CR spectrum when combined with SNR-CRs and EG-MPCS model (\citealt{Rachen2016}) for the extragalactic CRs. \textit{Bottom panel:} Mean logarithmic mass when combined with the EG-MPCS (red curve) and the EG-UFA (green curve, same as \ref{fig:log_mass} ) models. Data are the same as in Figure \ref{fig:log_mass}.} 
\label{fig:extragalactic_compare}
\end{figure}

As mentioned in \ref{section:2.2}, we consider two different models of extragalactic CRs:  UFA model \citep{Unger2015}  and a combination of PCS and Minimal model (MPCS model) \citep{Rachen2016, Thoudam2016}. Depending on the chosen extragalactic component, the value of injection fraction and maximum cutoff energy can slightly change. The UFA and MPCS models predict a significant contribution of extra-galactic cosmic rays below the `ankle'. 
All these different extragalactic models can explain the observations when combined with the SNR-CR component and the CR component from star clusters, although the UFA model somehow shows a smooth transition (Figure \ref{fig:all_comp}) between the Galactic and extragalactic components. The sharp increase near $10^9$ GeV in the MPCS model (top panel, figure \ref{fig:extragalactic_compare}) is due to the dip in the proton spectrum. It results from the intersection of the minimal model and the components from galaxy clusters. Below $10^9$ GeV, both the UFA and MPCS models give similar results and can explain the observed spectra.
We have also shown the mean logarithmic mass plot for a combination of each different extragalactic model with the two different Galactic components (bottom panel of figure \ref{fig:extragalactic_compare}). It is clear from the plot that all these different models for the extragalactic component, in combination with the Galactic components, follow the observed trend of mean logarithmic mass for the whole energy range. 
%
%
%
\section{Discussion}
\label{sec:discussion}
Our study demonstrates that the cosmic rays originating from spatially distributed young massive star clusters in the Galactic plane fit well the all-particle CR spectrum, particularly in the $10^7-10^9$ GeV energy range, and therefore this can be a potential candidate for the `second Galactic component' of CRs. We also show that the observed all-particle spectrum, as well as the cosmic ray composition at high energies, can be explained with the following 
two types of Galactic sources: (i) SNR-CRs,  dominating the spectrum up to $\sim 10^7$ GeV, and (ii) star cluster CRs, which dominate in the range $10^7-10^9$ GeV. 

The SNR-CR component can only contribute up to maximum energy, corresponding to a
proton cut-off energy of $2.5 \times 10^6$ GeV \citep{Thoudam2016}. Such a high value of energy cannot be achieved if we consider the DSA mechanism with typical values of the magnetic field in the ISM.  
However, some numerical simulations have indicated that supernova shocks can amplify the magnetic field near them several times larger than the value in the ISM \citep{Bell2001, Reville2012}. Such a strong magnetic field can accelerate CR protons up to the cut-off energy used in this study. Also, recently detected $\gamma$-rays from a few SNRs have also identified a few SNRs as cosmic ray PeVatrons that can accelerate particles up to a few times $\sim 10^6$ GeV energy.

{\em Maximum CR energy in star clusters:} 
{According to our model, the component of CRs that is plausibly generated in star clusters can contribute significantly towards the total CR flux, especially in the $10^7-10^9$ GeV range, if one considers that the protons can be accelerated up to $5\times 10^7$ GeV energy. For other elements with atomic number $Z$, the maximum energy is $5\times 10^7$ Z GeV in these young compact star clusters, and a cosmic ray injection fraction of $\sim 5\%$.
Note that this value of maximum energy required for proton is slightly on the higher side, but can be justified under the assumption of the very high initial shock velocity of SNRs inside compact clusters, faster wind velocity, and possible amplification of magnetic field inside the cluster. 
Our maximum CR energy from Hillas criterion may be an overestimate, 
but we should point
out that the 
requirement of $E_{\rm max} = 50$ PeV ($5\times 10^7$ GeV) depends on the assumption of elemental abundance ratios in the wind material, an aspect that remains uncertain at present. A higher abundance of heavy elements
would 
increase $E_{\rm max}$, as required to fit the observed spectrum. 
Also note that the recently detected $\gamma$-ray photons by {\it LHAASO} in PeV range from $12$ objects, some of which are associated with massive star clusters, have indicated that these clusters can accelerate particles at least up to a few tens of PeV \citep{Cao2021}, consistent with our estimates.  

{\em CR anisotropy:} The total CR anisotropy $\Delta$ can be calculated from our model using the diffusion approximation given by \citet{1972ChJPh..10...16M},
\begin{eqnarray}
    \Delta=\frac{3D}{c}\frac{|\nabla N|}{N}\,\,\,,
\end{eqnarray}
where $N$ is the CR number density, $D$ is the diffusion coefficient and $c$ is the velocity of light. From our model, we get $\Delta \sim 0.04-0.2$ in the range $1-100$ TeV. However, it is noteworthy that our calculated estimates exhibit higher values compared to the measured anisotropy, which is approximately in the range of $(0.5-1)\times 10^{-3}$ for the same energy spectrum. Notably, our findings align with the earlier estimates proposed by \citealt{2012JCAP...01..011B, 2012MNRAS.421.1209T}, although, like the case of previous calculations, they are larger than the observed anisotropy in the same energy range.

Inside an OB association, individual SNR shocks, as well as colliding shocks, can accelerate particles on a time scale below $1000$ years. An OB association may enter the evolutionary stage of multiple SN explosions on a time scale larger than a few hundred thousand years. It can create large bubbles of $\sim 50$ pc size, and the injected mechanical power can reach $\sim 10^{38}$ erg s$^{-1}$ over $10$ Myr—the lifetime of a superbubble. This process is supplemented by the formation of multiple shocks, large-scale flows, and broad spectra of MHD fluctuations in a tenuous plasma with frozen-in magnetic fields. The collective effect of multiple SNRs and strong winds of young massive stars in a superbubble is likely to energize CR particles up to hundreds of PeV in energy (see \citealt{Montmerle1979, Cesarsky1983, Bykov1992, Axford1994, Higdon1998, Bykov2001, Marcowith2006, Ferrand2010}) and even to extend the spectrum of accelerated particles to energies well beyond the `knee' \citep{Bykov2001}.

\cite{Gupta2020} pointed out the advantage of considering CRs accelerated in massive star clusters in explaining several phenomena (e.g., Neon isotope ratio) that are left unexplained by the paradigm of CR production in SNRs. They also proposed that this component need not be considered entirely independent and separate from the SNR component since massive stars (which are the progenitors of SNRs) always form in clusters. Therefore, the two components (SNR-CR and star cluster CRs) arise from similar sources, with some differences. The SNR-CRs can be thought of as CRs produced in individual SNRs, which arise from very small clusters with one or two massive stars, whereas the second component can be thought of as arising from different shocks that occur in the environment of massive star clusters. Hence, the two components can be put on the same platform, and the combined scenario offers a fuller, more complete picture of the phenomenon of CR acceleration in the Galaxy. 
 
\subsection{Caveats of our model}
Finally, we discuss some caveats of our model. 

{\em The injection fraction}, a free parameter in our analysis whose value is obtained by fitting the all-particle CR spectrum, ultimately depends on many other factors, such as diffusion coefficient, the assumed lower limit of the number of OB stars in a cluster, and so on. With a higher value of diffusion coefficient, the required injection fraction should be increased in order to match our results with observations. A larger diffusion coefficient implies that particles would diffuse out of the source more rapidly, which will decrease the particle density in the vicinity of clusters. This is why one needs a larger injection fraction to explain the observational data. On the other hand, the total number of OB associations depends on the lower cut-off in the distribution of cluster masses. E.g., the number of OB associations which has a minimum of 30 OB stars is lower than the number of OB associations with a minimum of 10 OB stars. For the second case, the required injection fraction will be lower (since the number of OB associations is higher). Also, the location of the peak of the cluster spatial distribution has a significant effect on the observed flux and may introduce some uncertainty to the value of the injection fraction parameter.

The efficiency is likely independent of the number of OB stars inside a single star cluster. Considering the gamma-ray luminosity of massive star clusters, observations indicate the gamma-ray luminosity is $\sim 10^{-3}\times L_w$ (where $L_w$ the wind kinetic energy), irrespective of the total Stellar number (e.g., \citealt{Ackermann2013}), as we have assumed here (see also \citealt{Gupta2018}). 
However, the required cosmic ray injection fraction to explain the observed data depends on the total number of star clusters in the Galaxy. For the current study, we have considered the spatial distribution of OB stars but we did not classify them according to their age and mass. According to the recent GAIA survey, $\sim 20\%$ of the total clusters are compact young, and massive \citep{Vieu2022}. However, we have assumed all the OB associations are young and compact. If we instead take a fraction (say $20\%$) of these clusters to be young then we would need a higher (by a factor of 5, say) fraction of cosmic ray injection efficiency to match the observed flux.

{\em Elemental abundance:} There are other uncertainties that arise from the assumed abundances of the eight elements considered here. This elemental abundance depends on the rotational velocity of the stars. For our calculations, we have used the abundances of stars, which rotate with a velocity that is $60\%$ of the critical velocity of the star. Varying the rotational velocity would change the elemental abundances. This will give an uncertainty between $2-3\%$ in the mean logarithmic mass plot (Figure \ref{fig:log_mass}). Results may also change if abundances from other previous works (\citealt{Heger2000, Heger2005}) were to be used. Comparing the abundances from these works with the ones used here, we find that it would introduce an uncertainty of $5-7\%$ in Figure 5. However, these variations will not significantly change the shape of our predicted $\langle \ln A \rangle$.

{\em CR propagation:} Another aspect that is important to mention is the mode of cosmic ray propagation. Our calculation of the second component 
assumes diffusion from the source. Nevertheless, it is crucial to acknowledge that the diffusion approximation may cease to be valid beyond a specific energy threshold, leading to a shift from diffusion to drift motion in the transport process. If we consider a mean magnetic field of $3$ $\mu$G in the Galactic plane then the transition from diffusion to drift will 
occur at $\sim Z \times 10^{17-18}$ eV \citep{Kaapa2023}. The maximum energy for proton in WTS is $\sim 50$ PeV ($5 \times 10^{16}$ eV) and is below the transition region. Therefore, the diffusion approximation works well for the energy range considered by us. However, we modified the diffusion coefficient above $10^{17}$ eV in a manner that mimics ballistic propagation beyond this energy threshold and found the spectra do not change significantly as the second component in this energy range is mainly dominated by the exponential cutoff.

\section{Conclusions}
\label{sec:conc}
In this paper, we suggest that the `second Galactic component of CRs', needed to explain the observed flux of CRs in the range between the `knee' and the `ankle' ($10^7$ GeV to $10^9$ GeV ), may arise from a distribution of massive star clusters. 
 
This component can bridge the gap between the SNR-CR component, which dominates below $\sim 10^7$ GeV, and the extragalactic component, which dominates above $\sim 10^9$ GeV. It has been previously noted that SNR-CRs and CRs from star clusters need not be considered two separate components, but rather originating from similar sources, {\it viz.} massive star clusters, the less massive ones leading to individual SNRs and SNR-CRs, while the more massive ones can accelerate CRs in a variety of strong shocks appearing in the dense cluster environment.  
We have argued that there is a possibility of acceleration of protons up to a few tens of PeV by considering the particle acceleration around the WTS, as well as acceleration by SNR shocks inside massive star clusters. This value is larger than that possible in the standard paradigm of CR acceleration inside supernova remnants present in the ISM. In this paper, we have carried out a detailed calculation of the propagation of cosmic rays in the Galaxy and demonstrated that this model can possibly explain the all-particle CR spectrum measured at the Earth. Our calculation considers a realistic distribution of star clusters in the Galaxy and also includes all the important transport processes of CRs including re-acceleration by the old SNRs in the Galaxy.

Our analysis requires a proton cut-off energy of $\sim 5 \times 10^7$ GeV ($50$ PeV) for the CRs accelerated in star clusters.  
A comparison of our analytical results with the observed all-particle CR spectrum yields an injection fraction (the fraction of kinetic energy of shocks being deposited in CRs) of $\sim 5\%$ (depending on the choice of the extragalactic component). 
Furthermore,  the variation of the mean logarithmic mass with CR energy (especially in the energy range of around $10^7\hbox{--}10^9$ GeV) supports the argument that the suggested CR component from star clusters can be considered as the second Galactic component of CRs.

\section*{Acknowledgements}
We would like to thank Arpita Roy for the stellar wind elemental abundance data and acknowledge the Australian supercomputer facilities GADI and AVATAR, where those simulations have been run. We thank Manami Roy, and Alankar Dutta for the valuable discussions. We also thank Thibault Vieu for the helpful discussions. SB acknowledges the Prime Minister's Research Fellowship (PMRF) and Govt. of India for financial support. ST acknowledges funding from the Abu Dhabi Award for Research Excellence (AARE19-224) and the Khalifa University Emerging Science \& Innovation Grant (ESIG-2023-008). PS acknowledges a Swarnajayanti Fellowship (DST/SJF/PSA-03/2016-17) and a National Supercomputing Mission (NSM) grant from the Department of Science and Technology, India.

\section*{Data Availability}
The data not explicitly presented in the paper will be available upon reasonable request from the first author.



\bibliography{sample631} 





\label{lastpage}
\end{document}